\documentclass{svjour3}

\usepackage{graphicx}
\usepackage{xcolor}
\usepackage{amsmath}
\usepackage{amssymb}
\usepackage{bm}

\newcommand{\vect}[1]{\mathbf{#1}}
\newcommand{\ud}{\ensuremath{\text{d}}}

\begin{document}

\title{Extracting vibrational anharmonicities from short driven molecular dynamics trajectories}
  
\author{Pascal Parneix \and Romain Maupin \and Loïse Attal \and Florent Calvo \and Cyril Falvo}

\institute{P. Parneix \at Université Paris-Saclay, CNRS, Institut des Sciences Moléculaires d’Orsay, 91405, Orsay, France \email{pascal.parneix@universite-paris-saclay.fr}
\and
    R. Maupin \at Université Paris-Saclay, CNRS, Institut des Sciences Moléculaires d’Orsay, 91405, Orsay, France
    \and
    L. Attal \at Université Paris-Saclay, CNRS, Institut des Sciences Moléculaires d’Orsay, 91405, Orsay, France
    \and
    F. Calvo \at University Grenoble Alpes, CNRS, LiPhy, 38000 Grenoble, France
    \and 
    C. Falvo \at Université Paris-Saclay, CNRS, Institut des Sciences Moléculaires d’Orsay, 91405, Orsay, France and University Grenoble Alpes, CNRS, LiPhy, 38000 Grenoble, France
}



\date{Received: date / Accepted: date}

\maketitle

\begin{abstract}
  Anharmonicities provide a wealth of information about the vibrational dynamics, mode coupling and energy transfer within a polyatomic system. In this contribution we show how driven molecular dynamics trajectories can be used to extract anharmonicity properties under very short times of a few hundreds of vibrational periods, using two exciting fields at- and slightly off-resonance. Detailed analyses on generic quartic potential energy surfaces and applications to various model systems are presented, giving good agreement with perturbation theory. Application to a realistic molecule, cubane (C$_8$H$_8$), modelled with a tight-binding quantum force field, further indicates how the method can be applied in practical cases.
\end{abstract}

\section{Introduction}

The harmonic approximation is ubiquitous in quantum chemistry and condensed matter physics as the starting point for vibrational or lattice dynamics. Besides providing numerically tractable treatments for high-dimensional problems, it also allows convenient representations of complex environments such as substrates or matrices with which generic systems of interest are in contact. In vibrational spectroscopy, and for molecules or coumpounds requiring accurate descriptions of electronic structure, the harmonic approximation for candidate geometries is usually the most straightforward approach for identifying experimentally measured infrared (IR) spectra. However, its performance depends on many factors, starting with the underlying electronic structure method itself, including the possible basis set uncompleteness, but also external elements such as a finite temperature or the existence of large-amplitude modes. Anharmonicities, which broadly define such effects lying beyond what the harmonic approximation can capture, are not only quantitative features but they are also essential to explain how vibrational modes are coupled with each other and how energy transfer is controlled. They can be experimentally accessed through various spectroscopic methods such as temperature-resolved infrared absorption measurements~\cite{joblin95,klingbeil07} or two-dimensional IR spectroscopy~\cite{Kim:2009qf,Hamm:2011cr,Falvo:2015ty}.

Anharmonicities in computed vibrational spectra can be mimicked by applying a scaling factor to the harmonic frequencies, which can often be in error by about 10--15\% relative to the measured frequencies~\cite{pople81,pulay83}. In addition to this empirical approach, anharmonic descriptions of vibrational dynamics are  captured exactly by solving the (nuclear) Schr\"odinger equation for the reference problem, but practical solutions are typically limited by the number of degrees of freedom and possibly the need to represent the multidimensional potential energy surface on a grid. Approximations to the nuclear dynamics, based e.g. on path integrals~\cite{pavese99,ramirez04,calvopimd10}, on semiclassical ideas~\cite{liu09,calvoqtb12}, or those resting on general approaches such as the Multimode code~\cite{bowman2010} or fixed-node diffusion Monte Carlo~\cite{Reynolds1982} provide other classes of methods that can be used to describe the anharmonic vibrational dynamics in classical or quantum sytems of reasonably large size. Anharmonic frequencies for quantum systems can also be obtained through dedicated methods such as principal component analyses or self-consistent phonons (quasiharmonic theory)~\cite{martinez06,calvoscp10}. Alternatively, second-order vibrational perturbation theory based on a quartic expansion of the potential energy surface~\cite{Mills:1972fu,Mulas:2018aa}, or VPT2, has become increasingly popular to determine accurate anharmonic corrections in IR spectra and, more recently, in vibrational circular dichroism~\cite{bloino12}.

The classical treatment of nuclear motion is yet another approximation that can contribute to addressing large systems. It is all the more accurate as temperature is high, in a regime where anharmonicities are also expected to be the strongest, and considerably simplifies the computational methodology with molecular dynamics (MD) simulations standing as a well-established tool for this purpose. In this context, Bowman and coworkers~\cite{Bowman2003,Kaledin2004,Kaledin:2006aa} proposed a method to extract the vibrational frequencies in large systems for which the solution of the harmonic problem would require diagonalizing excessively large Hessian matrices. The driven molecular dynamics (DMD) approach introduced by these authors adds an oscillatory driving external force to a molecular system, and evaluates its response as a function of the exciting frequency. The propensity of the driving field to heat the system as its frequency becomes resonant with a true vibrational mode provides an indirect way of accessing those modes. Obviously the accuracy of the DMD approach is limited by the resolution used in scanning the exciting oscillatory field, and it surely does not compete with exact diagonalization in very small systems. However, the method is appealing because it only requires short trajectories of a limited number of vibrational cycles, and it can be used only near an experimentally relevant frequency range.

Although it was initially intended to work at low excess energies, in a regime where the dynamics is nearly harmonic, it has not been used to extract information about anharmonicities themselves except in the work by Thauney {\em et al.}~\cite{Thaunay2015} where the authors used DMD as an assignment tool to identify effective (anharmonic) vibrational frequencies in molecules at finite internal energy. In the present contribution, we use DMD to extract the anharmonicities themselves, assuming the harmonic solution is known. The main idea of our approach is to carry two trajectories, at and slightly away from resonance, and to monitor the heating efficiency through the time evolution of the internal energy, which due to anharmonicities also has an oscillatory character. In particular, as suggested by a rigorous analysis on a one-dimensional model system, the maximum energies reached by the system at these first oscillations and their comparison provide the magnitude and the sign of the anharmonic coefficient, respectively. More generally, in a many-body system, the method can be applied to yield the intra- and inter-mode contributions to anharmonicity and the entire set of couplings leading to the Dunham matrix. Application to the realistic case of the highly symmetric cubane molecule (C$_8$H$_8$), modeled here using a tight-binding quantum potential energy surface, indicates that the DMD method can produce anharmonic coefficients of comparable accuracy as the VPT2 approach, suggesting a significantly reduced computational cost for large molecular systems.

The article is organized as follows. In the next section, we lay out the generic strategy and explain, on a one-dimensional quartic potential, how the anharmonicities can be numerically obtained from the time evolution of the internal energy in DMD trajectories. Extension to coupled oscillators is then discussed and illustrated on other model systems, before turning to the application to cubane. Finally we give some concluding remarks in Sec. \ref{sec:ccl}.




\section{Theory and Methods}
\label{sec:methodology}

\subsection{Vibrational Hamiltonian}

We consider the classical vibrational dynamics of a $N$-atom molecular system, assuming the total linear and angular momenta are both set to zero. The dynamics can be described in terms of $g=3N-6$ normal mode coordinates collectively denoted as $\vect{q}=(q_1,\dots,q_g)$ together with their corresponding momenta $\vect{p}=(p_1,\dots,p_g)$.
Without any loss in generality, the Hamiltonian of the system is written as
\begin{equation}
H_0(\vect{q},\vect{p}) = \sum_i \frac{1}{2} p_i^2 + \frac{1}{2}\omega_i^2 q_i^2 + V_a(\vect{q}),
\label{eq:H0}
\end{equation}
where $\omega_i$ are the harmonic frequencies and $V_a(\vect{q})$ is the anharmonic contribution to the potential energy surface. In the context of canonical second-order perturbation theory, $V_a$ can be expanded up to 4th order in the normal modes coordinates $\vect{q}$~\cite{Mills:1972fu}, resulting in
\begin{equation}
V_a(\vect{q}) = \frac{1}{3!}\sum_{ijk}\Phi_{ijk} q_i q_j q_k + \frac{1}{4!}\sum_{ijk\ell}\Phi_{ijk\ell} q_i q_j q_k q_\ell,
\label{eq:quarticpotential}
\end{equation}
where $\Phi_{ijk}$ and $\Phi_{ijk\ell}$ are the cubic and quartic derivatives of the potential, respectively. Still within canonical perturbation theory, we introduce the harmonic angle-action variables~\cite{Goldstein:2000aa} $\{ I_i, \theta_i \}$ such that
\begin{align}
q_i &=- \sqrt{\frac{2I_i}{\omega_i}}\cos\left(\theta_i\right),\label{eq:q_angleaction_harm} \\
p_i &= \sqrt{2I_i\omega_i}\sin\left(\theta_i\right).
\end{align}
Using second-order canonical perturbation theory based on a perturbative canonical transformation~\cite{Goldstein:2000aa,Schatz:1979aa,Fried:1987aa,Sugny:1999aa}, we obtain the perturbative Hamiltonian as
\begin{equation}
H_0(\vect{I}) = \sum_i \omega_i I_i + \sum_{i\leqslant j} \chi_{ij} I_i I_j,
\label{eq:Haction}
\end{equation}
where ${\bf I}=(I_1, \dots,I_g)$ and $\chi_{ij}$ are the Dunham coefficients~\cite{Mills:1972fu} expressed as 
\begin{align}
\chi_{ii} &= \frac{1}{16} \frac{ \Phi_{iiii}}{\omega_i^2} \, - \frac{1}{16}\sum_{k} \frac{ \Phi_{iik}^2}{\omega_i^2\, \omega_k^2} \, \frac{ 8 \omega_i^2 - 3 \omega_k^2}{4 \omega_i^2 - \omega_k^2}    
\label{eq:dunham1} \\
 \chi_{ij} &= \frac{ 1}{4} \,  \frac{ \Phi_{iijj}}{\omega_i  \, \omega_j} -    \frac{ 1}{4} \, \sum_{k} \frac{ \Phi_{iik} \, \Phi_{kjj}}{\omega_i  \, \omega_j \, \omega_k^2}    -    \frac{ 1}{2} \, \sum_{k} \frac{ \Phi_{ijk}^2}{\omega_i  \, \omega_j \, \Delta_{ijk} } \, (\omega_k^2-\omega_i^2-\omega_j^2) 
\label{eq:dunham2}
\end{align}
and which involve the coefficients $\Delta_{ijk}$ defined as
 \begin{equation}
\Delta_{ijk} = (\omega_i -\omega_j -\omega_k) \,    (\omega_i +\omega_j -\omega_k) \,  (\omega_i -\omega_j +\omega_k) \,    (\omega_i +\omega_j +\omega_k).
\label{eq:delta}
\end{equation}
Note that in the quantum-mechanical context, and due to the specific ordering of the operators, an additional constant that modifies the zero-point energy needs to be accounted for in Eq.~(\ref{eq:Haction})~\cite{Schatz:1979aa}.
The entire set of Dunham coefficients fully determines the anharmonicities, and in particular their effect on the vibrational frequencies, the harmonic values $\omega_i$ being replaced by their anharmonic counterparts $\tilde \omega_i$ as
\begin{equation}
\tilde{\omega}_i\left(\vect{I}\right) = \frac{\partial H}{\partial I_i} = \omega_i + 2\chi_{ii}I_i + \sum_{j\neq i} \chi_{ij} I_j.
\end{equation}
The latter equation notably expresses that the anharmonic frequencies depend on the total energy via the action variables. Assuming energy equipartition  between the modes, the actions can be estimated in the harmonic approximation as a function of the total energy as $I_i = E/(g \omega_i)$, which allows the anharmonic frequencies to be rewritten explicitly as
\begin{equation}
\tilde{\omega}_i\left(E\right) = \omega_i\left( 1 + \gamma_i E \right),
\end{equation}
where the anharmonic coefficients $\gamma_i$ are related to the Dunham coefficients through
\begin{eqnarray}
\gamma_i &=& \gamma_i^{\rm intra} + \gamma_i^{\rm inter} \label{eq:gtotal} \\
\gamma_i^{\rm intra} &=& \frac{2\chi_{ii}}{g\omega_i^2} \label{eq:gintra} \\
\gamma_i^{\rm inter} &=& \sum_{j\neq i} \frac{\chi_{ij}}{g\omega_j\omega_i},
\label{eq:ginter}
\end{eqnarray}
and were the intramode and intermode contributions $\gamma_i^{\rm intra}$ and $\gamma_i^{\rm intra}$ to the total anharmonicity of mode $i$ have been introduced. 

In the following, we focus on the determination of the anharmonic coefficients $\gamma_i$ and their two components using driven molecular dynamics but our method can be adapted to calculate the entire Dunham matrix. 

\subsection{Driven molecular dynamics}
\label{ssec:mol}

We now exploit a particular type of molecular dynamics scheme to infer the anharmonic coefficients from the response of the system to a time-dependent perturbation, as originally introduced by Bowman and coworkers for the harmonic problem \cite{Bowman2003}. More precisely, the system is coupled to an external field oscillating at the angular velocity $\Omega$, and the coupling generally reads 
\begin{equation}
H_c(\vect{q},t;\Omega) = -F(\vect{q}) \sin\left(\Omega t\right),
\end{equation}
where in a realistic interaction model between the molecule and the electromagnetic field  $\vect{E}$ the coupling would read $F(\vect{q}) = \bm{\mu}(\vect{q}) \cdot \vect{E}$, with $\bm{\mu}$ the electric dipole moment vector.

Without loss of generality, and assuming the coupling with the external field is low in magnitude, we can expand $F(\vect{q})$ linearly in the coordinates as $F(\vect{q}) = \sum_i F_i q_i$, leading to the total Hamiltonian 
\begin{equation}
H(\vect{q},\vect{p},t) = H_0 (\vect{q},\vect{p}) - \sum_i F_i q_i \sin\left(\Omega t\right).
\end{equation}
For a given excitation frequency close to a specific natural frequency of the system, the total energy will increase due to the coupling with the field. However, due to anharmonicities, the energy itself will acquire an oscillatory character and reach a first maximum that we denote as $\overline{E}$. This quantity will precisely provide the connection needed with the anharmonic coupling coefficients $\gamma_i$ that we seek to determine.
\subsection{One-dimensional system}
The case of one-dimensional systems is particularly insightful as it allows the phenomenology to be explored and the quantitative connections between the response of the driven system and the anharmonic coefficients to be established. For $g=1$ the Hamiltonian simplifies as
\begin{equation}
H(q,p,t) = \frac{p^2}{2} + \frac{1}{2}\omega^2 q^2 + \frac{1}{3!}\phi^{(3)} q^3 + \frac{1}{4!}\phi^{(4)} q^4 - F q \sin(\Omega t),
\label{eq:1D_Hmodel}
\end{equation}
and the problem corresponds to the Duffing oscillator which has been extensively studied in the literature~\cite{Struble1962,Hale1969,Hagedorn1988,Jordan1999}. It is well known that, for a resonant excitation, the energy of an harmonic oscillator increases with time as $\propto t^2$. In the case of the Duffing model, the anharmonicity originating from the cubic $\phi^{(3)}$ and quartic $\phi^{(4)}$ coefficients induces a dephasing of the oscillator with respect to the excitation force, which in turn results in oscillations in the energy. Here and in the following, resonances have to be understood relative to the harmonic limit, rather than the true energy- or temperature-dependent vibrational frequency that incorporates anharmonic effects.

It is possible to relate the maximum energy reached by the system to the anharmonicity coefficients, and in this goal we use second-order perturbation theory and the angle-action variables to rewrite the above Hamiltonian  as
\begin{equation}
H(\theta,I,t) = \omega I +\chi  I^2 + \frac{F}{\sqrt{\omega}}\sqrt{2I} \cos(\theta)\sin\left(\Omega t\right),
\label{eq:H1d_a}
\end{equation}
where the single anharmonic parameter $\chi$ is given by
Eq.~(\ref{eq:dunham1}) or
\begin{equation*}
\chi = \frac{1}{16}\frac{\phi^{(4)}}{\omega^2} - \frac{5}{48}\frac{\left(\phi^{(3)}\right)^2}{\omega^4}.
\end{equation*}
In Eq.~(\ref{eq:H1d_a}), the coordinate $q$ in the coupling has been replaced by its harmonic expression given by Eq.~(\ref{eq:q_angleaction_harm}). In principle, other terms arising from the perturbative expansion in the anharmonic couplings should be included, but for weak enough couplings with the field we safely neglect these terms.

The Hamiltonian of Eq.~(\ref{eq:H1d_a}) can be simplified if we use the canonical transformation   $\left(\theta,I\right)\rightarrow \left(\psi,J\right)$ defined by
\begin{equation}
\begin{split}
\psi &= \theta - \Omega t, \\
J &= I.
\end{split}
\end{equation}
This time-dependent canonical transformation is equivalent to the Van der Pol transformation usually introduced to treat the Duffing oscillator with the method of averaging~\cite{Hale1969}. It derives from the generating function~\cite{Goldstein:2000aa}
$F(\theta, J ) = \left(\theta - \Omega t \right) J$,
from which the new Hamiltonian $K(\psi,J)$ reads
\begin{equation}
K(\psi,J,t)= H + \frac{\partial F}{\partial t} = - \delta J +   \chi J^2   -   \alpha \sqrt{2J} \sin \psi +    \alpha \sqrt{2J} \sin\left(\psi+2\Omega t \right) ,
\label{eq:HK1}
\end{equation}
with $\delta=\Omega-\omega$ and $\alpha=F/(2\sqrt{\omega})$. In Eq.~(\ref{eq:HK1}), if the coupling $\alpha$ is small enough, the last term oscillates rapidly and only has a weak effect on the system, it can therefore be neglected. This approximation is equivalent to the method of averaging usually performed on the equations of motion~\cite{Hale1969,Hagedorn1988}. The approximate Hamiltonian then becomes
\begin{equation}
K(\psi,J) = - \delta J +   \chi J^2   -   \alpha \sqrt{2J} \sin \psi.
\label{eq:HK2}
\end{equation}
Introducing now dimensionless variables,
\begin{align}
a &= \left( \frac{\chi}{\alpha} \right)^{1/3} \sqrt{2J}, \\
\tau &= \left(\alpha^2 \chi\right)^{1/3} t, \\
\epsilon &= \left(\alpha^2 \chi \right)^{-1/3} \delta,\label{eq:epsilon}
\end{align}
the Hamiltonian is rewritten as
\begin{equation}
K =\left(\frac{\chi}{\alpha^4}\right)^{1/3}\left(-\frac{\epsilon}{2} a^2 + \frac{1}{4} a^4 - a\sin\psi\right),
\label{eq:Kvalue}
\end{equation}
and remains a conserved quantity, although not equivalent to the system energy. To obtain the energy, it is necessary to go back to the original phase-space representation, which for our dimensionless variables yields
\begin{equation}
E = K - \frac{\partial F}{\partial t}
 = K + \frac{\Omega}{2} \left(\frac{\alpha}{\chi}\right)^{2/3} a^2.
 \label{eq:energydef}
\end{equation}
The time dependence of the energy can be inferred from solving the equations of motion that are straightforwardly obtained using Hamilton's equations derived from the Hamiltonian $K(\psi,J)$ Eq.~(\ref{eq:HK2}).
\begin{align}
& \frac{\ud a}{\ud \tau} = \cos\psi , \label{eq:eom1} \\
& a \frac{\ud \psi}{\ud \tau} = -\sin\psi - \left(\epsilon - a^2\right) a.
\label{eq:eom2}
\end{align}
Eq.~(\ref{eq:energydef}) shows that the energy is maximum when $a$ itself is, a condition that corresponds to $\overline{\psi}=\pm \frac{\pi}{2}$, as shown by Eq.~(\ref{eq:eom1}). In addition, if we now consider the specific initial conditions $(q,p)=(0,0)$ for which $K=0$, the maximum amplitude $\overline{a}$ is obtained by solving
\begin{equation}
\overline{a}^3 - 2\epsilon \overline{a} - 4 = 0.
\label{eq:amaxeq}
\end{equation}
Near resonance ($\Omega \simeq \omega)$, $\epsilon\ll 1$ and the maximum amplitude is 
\begin{equation}
    \overline{a} = 2^{2/3} + \frac{2^{1/3}}{3} \epsilon + \mathcal{O}\left(\epsilon^2\right),
\end{equation}
leading to the maximum in the energy given by
\begin{equation}
    \overline{E} = \Omega \left(\frac{\alpha}{\chi}\right)^{2/3} \left( 2^{1/3} + \frac{2}{3}\epsilon \right) + \mathcal{O}\left(\epsilon^2\right).
    \label{eq:Ebardelta}
\end{equation}
Exactly at resonance $\epsilon=0$, the maximum of the energy can thus be expressed as a function of the anharmonicity $\gamma=2\chi/\omega^2$ through
\begin{equation}
\overline{E}_0 = \overline{E}(\delta=0)=\left(\frac{\sqrt{2} F}{\omega\gamma}\right)^{2/3}.
\label{eq:gamma1d}
\end{equation}
This equation provides a means to determine the modulus $|\gamma|$ of the anharmonicity coefficient from the time variations of the total energy under the external driving force. However, owing to the presence of the power 2/3 in the above equation the sign of $\gamma$ remains undetermined at this stage.

To identify the sign of $\gamma$, an additional DMD trajectory is carried out, slightly away from resonance (nonzero $\delta$). Still assuming a weak coupling $F\ll \omega / \sqrt{|\gamma|}$, the maximum of the energy is now given by
\begin{equation}
\overline{E}_\delta \approx \overline{E}_0 + \frac{4 \delta}{3\gamma \omega},
\label{eq:emaxdelta}
\end{equation}
which by comparison with $\overline{E}_0$ yields the sign of $\gamma$ for a fixed $\delta$. In particular, for $\delta>0$ we expect the energy to increase relative to the resonance case if $\gamma>0$, and to decrease in the opposite situation.

The time $\overline{T}$ at which the energy reaches its first maximum can also be determined for the one-dimensional problem and the same initial conditions $(q=0,p=0)$ corresponding to $K=0$. This quantity is relevant in practice to evaluate the computational effort associated with the integration of the equations of motion of the DMD simulations. At resonance, and using Eqs.~(\ref{eq:Kvalue}) and~(\ref{eq:eom1}) we obtain 
\begin{equation}
\overline{T}_0 = \left(\alpha^2 |\chi|\right)^{-1/3} \overline{\tau}_0
\end{equation}
where $\overline{\tau}_0$ is given by
\begin{equation}
\overline{\tau}_0 = \int_{0}^{2^{2/3}} \frac{\ud a}{\sqrt{1-a^6/16}} = 2^{2/3} \sqrt{\pi}\frac{\Gamma(7/6)}{\Gamma(2/3)} \approx 1.92762,
\end{equation}
leading to
\begin{equation}
\overline{T}_0 \approx 3.85524 \left( F^2 \omega |\gamma| \right)^{-1/3}.
\label{eq:time}
\end{equation}
From $\overline{T}_0$, the modulus of the anharmonicity coefficient can also be estimated and compared against the value obtained from the maximum energy  $\overline{E}_0$. 
%
%

\subsection{$g$-dimensional system}

Aiming to extend the previously described approach to the general case of a $g$-dimensional system with $g>1$, we aim to evaluate the individual anharmonicity coefficients $\gamma_i$ defined by Eq.~(\ref{eq:gtotal}). Given the freedom in choosing the external field, we assume that the driving force only acts on a single mode $i$, all other modes being initially excited with a fixed constant energy $E_j=E_{\rm init}$ for all $j\ne i$.

The intramode anharmonicity $\gamma_i^{\rm intra}$ can be determined using the protocol described in the previous section, without injecting any excess energy in other modes, i.e. taking $E_{\textrm{init}}=0$. 

To calculate the intermode anharmonicity $\gamma_i^{\rm inter}$, some excess energy must be deposited in the other modes. We start again within the context of second-order perturbation theory, from which it can be shown that the actions $I_j$ for $j\neq i$ are all constants and their values can be approximated by $I_j=E_{\textrm{init}}/\omega_j$. Dropping the index $i$ for the sake of simplifying notations, the Hamiltonian describing the dynamics of the specific mode $i$ is then given by 
\begin{equation}
H(\theta,I,t) = \tilde{\omega}(E_{\textrm{init}}) I +\chi I^2 + \frac{F}{\sqrt{\omega}}\sqrt{2I} \cos(\theta)\sin\left(\Omega t\right),
\end{equation}
where $\tilde \omega(E_{\textrm{init}})$ is the modified frequency of mode $i$ due to the couplings with the other modes:
\begin{equation}
 \tilde{\omega}(E_{\textrm{init}}) = \omega \left( 1+ \gamma^{\text{inter}} g E_{\textrm{init}}\right).
\end{equation}
Following the same procedure as in the previous section, the Van der Pol transformation leads to the following time-independent Hamiltonian
\begin{equation}
K(\psi,J) = -\delta(E_{\text{init}}) J + \chi J^2 -\alpha\sqrt{2J}\sin\psi
\end{equation}
where the detuning $\delta$ is now given by
\begin{equation}
\delta = \Omega - \omega - \omega \gamma^{\text{inter}} g E_{\text{init}}.
\label{eq:deltabeta}
\end{equation}
As in the one-dimensional case, we first perform a DMD simulation exactly at resonance ($\Omega=\omega$) and taking $(q,p)=(0,0)$ again corresponding to $K=0$. Using Eq.~(\ref{eq:energydef}) we obtain the maximum energy reached by mode $i$ under scrutiny as
\begin{equation}
\overline{E}(E_{\text{init}}) = \frac{\omega}{2} \left( \frac{\alpha}{\chi} \right)^{2/3} \overline{a}^2(E_{\text{init}}).
\label{eq:Emaxval}
\end{equation}
Choosing $E_{\textrm{init}}=\overline{E}/2$, using 
Eqs.~(\ref{eq:epsilon}) and (\ref{eq:deltabeta}) and the fact that $\gamma^{\rm intra}=2\chi/(g\omega^2)$, we obtain the dimensionless detuning $\epsilon$ as
\begin{equation}
\epsilon = -\frac{\gamma^{\textrm{inter}}}{2\gamma^{\textrm{intra}}} \overline{a}^2.
\end{equation}
Using Eq.~(\ref{eq:amaxeq}) we finally obtain the maximum energy
\begin{equation}
\overline{E} = \left(\frac{\sqrt{2}F}{g \omega\left( \gamma^{\textrm{intra}} + \gamma^{\textrm{inter}}\right)}\right)^{2/3}.
\label{eq:Ebargdim}
\end{equation}
The anharmonicity inferred this way requires $E_{\rm init}=\overline{E}/2$, but this condition is not necessarily satisfied because the maximum energy $\overline{E}$ reached upon the DMD trajectory is not known a priori. This issue can be alleviated by conducting a series of DMD simulations at increasing excess energies $E_{\rm init}$, or alternatively as a function of the ratio $\eta$ defined by
 \begin{equation}
\eta=\frac{\sum_j^g E_j}{E_i}
\end{equation}
in which $E_j$ is the harmonic energy in the $j^{\rm{th}}$ vibrational mode, the condition $E_{\textrm{init}} = \overline{E}/2$ being then equivalent to $\eta=(g+1)/2$. 

In practice, effective anharmonicities $\gamma_i(\eta)$ are obtained as a function of $\eta$ from the DMD trajectories by using Eq.~(\ref{eq:Ebargdim}) and extrapolated if necessary for $\eta$ to reach $(g+1)/2$. The true anharmonicity $\gamma_i$ for mode $i$ is then taken at this value of $\eta$.

\section{Results and discussion}

The above method is now applied to a variety of practical situations, starting with systematic 1D and many-dimensional problems, but treating a more realistic molecular system in subsection \ref{ssec:cubane}. In all numerical applications, the molecular dynamics trajectories were integrated using the velocity Verlet method.

\subsection{1D quartic potential}
\label{ssec:1Dqp}
\begin{figure}
\centering
 \includegraphics[width=8.cm]{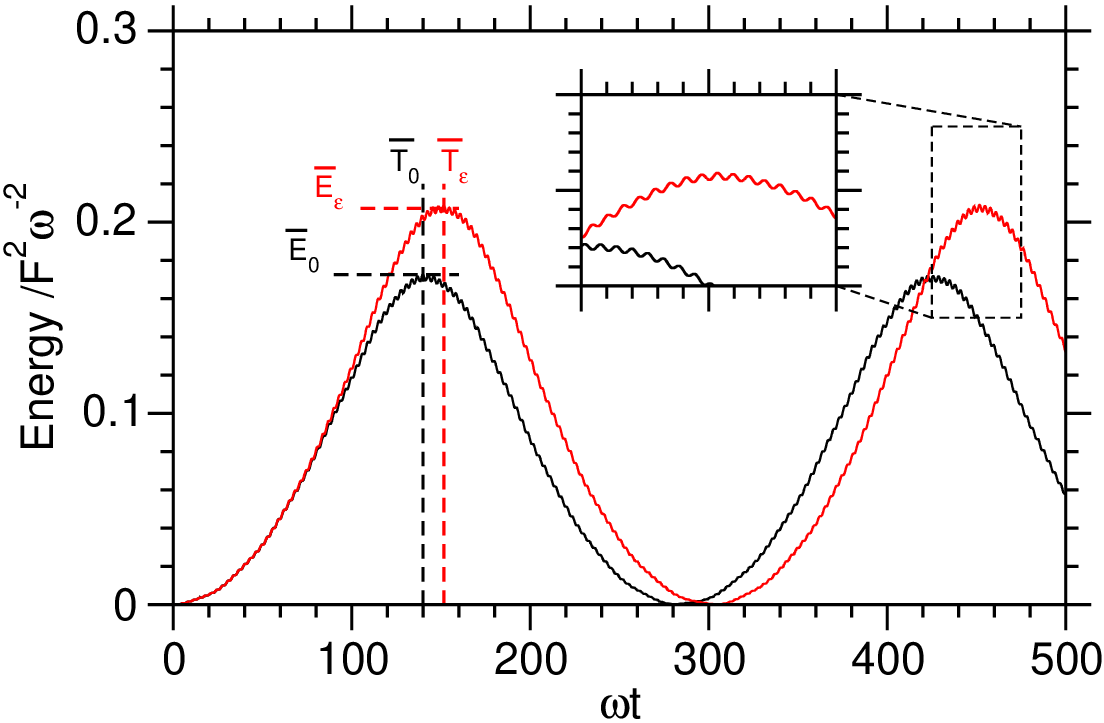}
 \caption{Energy as a function of time along two DMD trajectories with $F$=0.01, $\omega$ = 1 and ($\phi^{(3)}=0.3,\phi^{(4)}=1.8$). The black curve corresponds to a simulation at resonance ($\delta = 0$) and the red curve corresponds to a simulation with $\delta$ = 5 10$^{-3} \times \omega$. The inset is a closeup of a particular region highlighting the lower intensity signal at higher frequency $2\Omega$.} 
 \label{fig:fig1}
\end{figure}
We first consider the one-dimensional problem whose dynamics is governed by the Hamiltonian of Eq.~(\ref{eq:1D_Hmodel}), and analyse the evolution of the total energy as a function of time. For this sytem, all DMD simulations were initialized with $(q,p)=(0,0)$ at time $t=0$ and were integrated with a time step of 0.005 reduced units.

Fig.~\ref{fig:fig1} shows the particular evolution of the total energy obtained for specific choices in the anharmonic parameters and the external force. As expected, these variations exhibit a first maximum $\overline{E}_0$ at a time $\overline{T}_0$, and continue oscillating at longer times. As emphasized in the inset of Fig.~\ref{fig:fig1}, a smaller oscillation can be discerned on top of the main signal, which is a manifestation of the last oscillating term in Eq.~(\ref{eq:HK1}) at the higher frequency $2\Omega$. The noise arising from the mere presence of this extra oscillation can be ignored by considering only the main envelope of the signal when determining the maximum energy $\overline{E}$.

Repeating the DMD trajectory slightly off resonance ($\delta=0.005 \times \omega>0$) leads to a qualitatively similar behavior in the total energy of the system, but the maximum is shifted to higher values and delayed to a longer time, both features suggestive of a positive anharmonic coefficient for this specific system.


%
\begin{figure}
\centering
\includegraphics[width=8.cm]{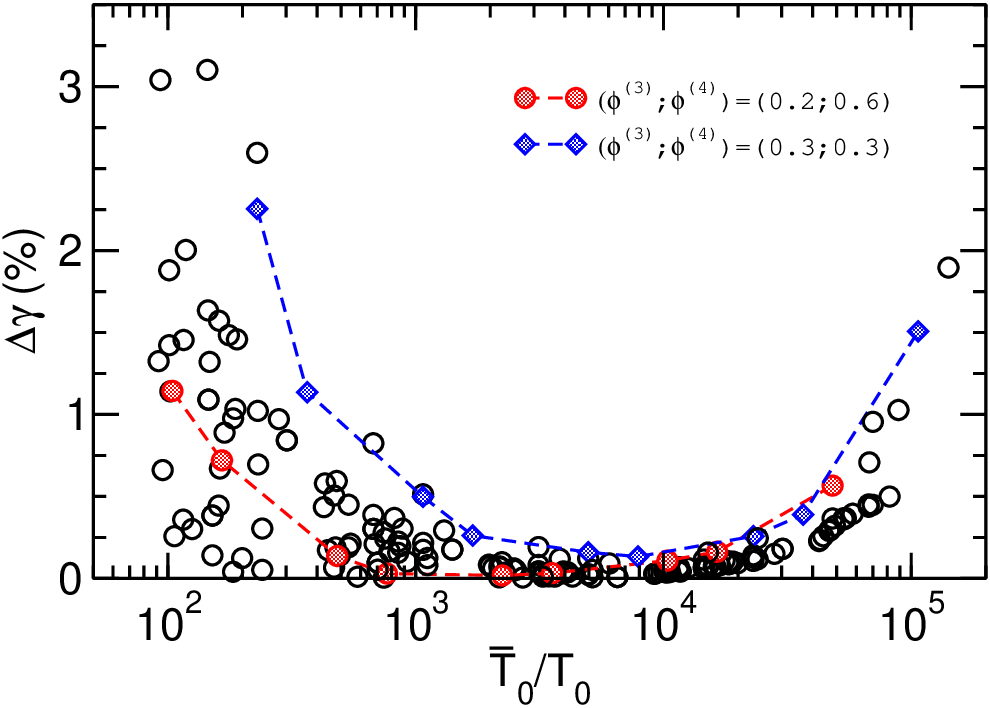}
  \caption{Relative error in the prediction of the anharmonic coefficient $\gamma$ with respect to the reference value from perturbation theory, as a function of the normalized time of the energy maximum  $\overline{T}_0/T_0$ obtained using 9 different driving forces $F$. The results are shown for 18 systems, with DMD simulations performed with $\delta=5 \, 10^{-6} \times \omega$. For two particular systems, the values of the error for increasing force $F$ are connected by blue or red lines to guide the eye.}
 \label{fig:fig2}
\end{figure}

Such simulations were carried out for 18 independent anharmonic oscillators characterized by different values of the $\phi^{(3)}$ and $\phi^{(4)}$ parameters and using 9 values of the external force $F$ ranging from 10$^{-7}$ up to 10$^{-3}$ for each set of parameters. Keeping a harmonic frequency of $\omega=1$, the cubic and quartic coefficients were taken randomly with values between -1 and +2 so as to yield reasonable values, either positive or negative, for the anharmonic coefficients.

The anharmonic coefficient $\gamma$ was determined from the above described method, first carrying a DMD trajectory at resonance, and another one slightly off resonance. The value $\gamma_{\rm DMD}$ obtained this way was compared with that ($\gamma_{\rm PT}$) predicted by perturbation theory, Eq.~(\ref{eq:dunham1}), by calculating the relative error $\Delta \gamma$ between these two values.
The various errors $\Delta \gamma$ are shown in Fig.~\ref{fig:fig2} as a function of the time $\overline{T}_0$ at which the energy reaches a maximum, normalized by the period $T_0=2\pi/\omega$ of the harmonic oscillator.

Overall, this error is always lower than a few percents in the entire range of explored parameters and external driving forces.
Interestingly, the error is always minimum when $\overline{T}_0$ is about a few hundred vibrational periods, but increases away from this approximate range. As shown in Eq.~(\ref{eq:time}), $\overline{T}_0$ is proportional to $F^{-2/3}$. If the driving force becomes too large, then $\overline{T}_0$ can be very short but  the dynamics may no longer be described in the perturbative regime. Conversely, if the driving force is very small, then $\overline{T}_0$ becomes very long and the integration of the DMD trajectory will suffer from numerical integration errors. In both cases, the accuracy in determining the anharmonic coefficient drops, which is the trend observed in Fig.~\ref{fig:fig2}.

From this phenomenological study on model 1D systems, we conclude that the DMD method is quantitative for estimating anharmonicities, and is best conducted over durations of a few hundred vibrational periods. 


\subsection{Coupled anharmonic potential}

The method was next applied to model 5-dimensional coupled anharmonic systems described by the quartic potential of Eq.~(\ref{eq:quarticpotential}). The cubic and quartic parameters employed for each of the 5 models were chosen randomly in the range between -0.7 and +1.5 in order to yield reasonable anharmonic coefficients, positive or negative. As was the case for the 1D system previously discussed, perturbation theory is exact for such systems in providing the anharmonicity coefficients and, for the present multidimensional problem, their intramode and intermode contributions.

Following the procedure laid out in the previous section, DMD simulations were carried out by exciting each mode $i$ one mode at a time, i.e. taking $F_i=10^{-5}$ and $F_j=0$ for $j\neq i$, and a detuning $\delta=5\cdot 10^{-6}\times \omega_i$. No excess energy was added to the mode under scrutiny, but some initial energy $E_{\rm init}$ was deposited randomly into each other mode, which we measure through the parameter $\eta$ describing the relative amount of harmonic energy into the spectator modes. After monitoring the time variations of the total energy of the system, the anharmonicity $\gamma_i(\eta)$ was then computed using Eq.~(\ref{eq:Ebargdim}). The process was reiterated by varying $E_{\rm init}$ or $\eta$, the true anharmonicity $\gamma_i$ of mode $i$ being taken at $\eta=(g+1)/2=3$. For this system, the integration step was again of 0.005 reduced time units.

\begin{figure}
\centering
 \includegraphics[width=8.cm]{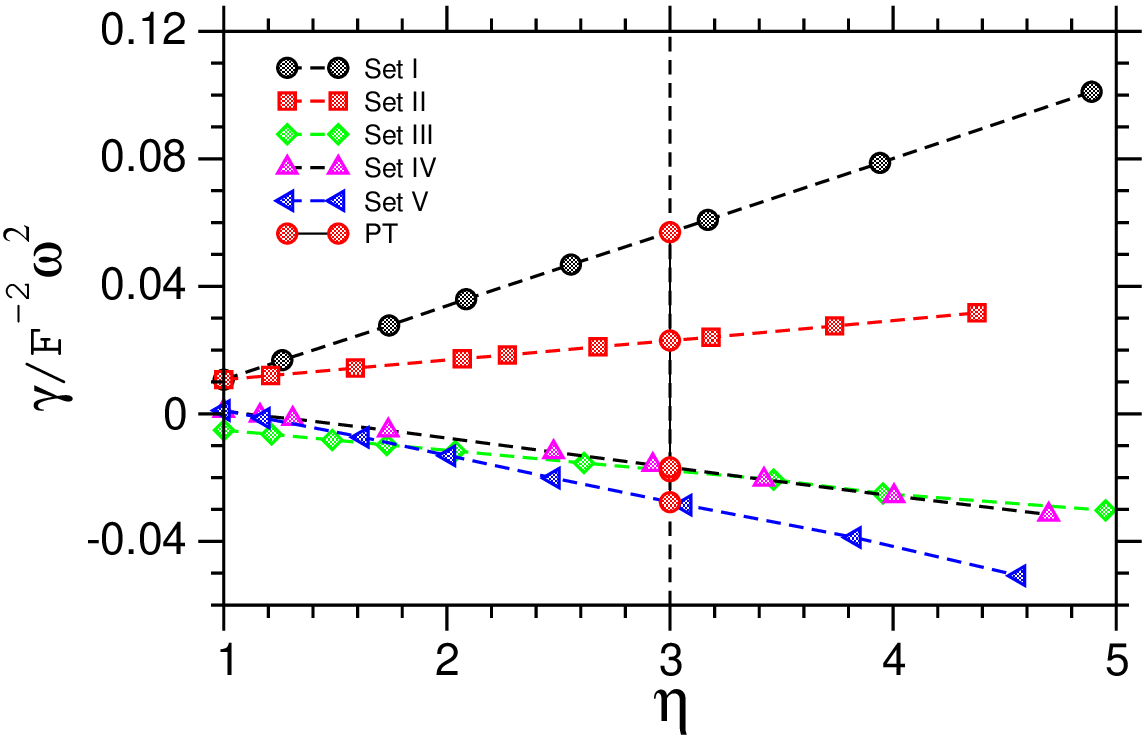}
  \caption{Effective anharmonicity $\gamma$ obtained for the first mode of coupled 5-mode systems using DMD trajectories as a function of the relative excess energy $\eta$ deposited in the spectator modes. The results are shown for five independent systems. The true anharmonicities expected for $\eta=3$ are highlighted by their intersection with the vertical dashed line. The values of the anharmonicities predicted by perturbation theory are also given on this line.}
 \label{fig:fig4}
\end{figure}

Dropping the index $i$, we show in Fig.~\ref{fig:fig4} how the effective anharmonicity $\gamma$ of one specific mode varies with increasing relative excess energy $\eta$ deposited in the other modes, for the five models.
The effective anharmonicity appears to vary essentially linearly with increasing $\eta$, which allows a straightforward determination of the true anharmonicity at $\eta=3$. Such a linear behavior was expected from equation (\ref{eq:emaxdelta}) and the linear dependence of the detuning with the excess energy expressed by Eq.~(\ref{eq:deltabeta}).

The intramode contribution $\gamma^{\rm intra}$ to $\gamma$ can be estimated independently, by performing an additional DMD simulation starting without any excess energy but exciting the mode of interest. The intermode contribution $\gamma^{\rm inter}$ is then obtained by simple difference from $\gamma$. The resulting values for the intramode and intermode contributions obtained from DMD trajectories and from perturbation theory are compared with each other in Table \ref{tab:2DD} for the five systems.

\begin{table}
\begin{center}
\begin{tabular}{c|rrr|rrr|r}
\hline \hline
Parameters & \multicolumn{3}{c|}{DMD} &
\multicolumn{3}{c|}{PT} & Error \\
set & intra & inter & total & intra & inter & total &
(\%) \\
   \hline
   I    &  0.0107 &  0.0462 &  0.0569 & 0.0108 & 0.0462  &  0.0570 & 0.2\\
   II   &  0.0107 &  0.0123 &  0.0230 & 0.0108 & 0.0122 &  0.0230 & 0.0\\
   III  & -0.0052 & -0.0126 & -0.0178 & -0.0045 & -0.0134  & -0.0180 & 1.1\\
   IV   &  0.0009 & -0.0174 & -0.0165 & 0.0016 & -0.0184 & -0.0168 & 1.8\\
   V    &  0.0010 & -0.0286 & -0.0276 & 0.0017  & -0.0294 & -0.0277 & 0.4\\
   \hline\hline

\end{tabular}

\caption{Anharmonicity coefficients $\gamma$ and their intramode and intermode contributions $\gamma^{\rm intra}$ and $\gamma^{\rm inter}$ computed from DMD simulations and given by perturbation theory, for the five coupled oscillator systems. The relative error is given for the total anharmonicity, in percents.}
\label{tab:2DD}
\end{center}
\end{table}


Overall the agreement between the two methods is again very satisfactory, the maximum error amounting to less than 2\%. However, it is fair to recognize that some compensation in the errors associated to the intramode and intermode contributions occured, intramode (intermode) anharmonicities being generally underestimated (overestimated) by the DMD method.

\begin{figure}
\centering
 \includegraphics[width=8.cm]{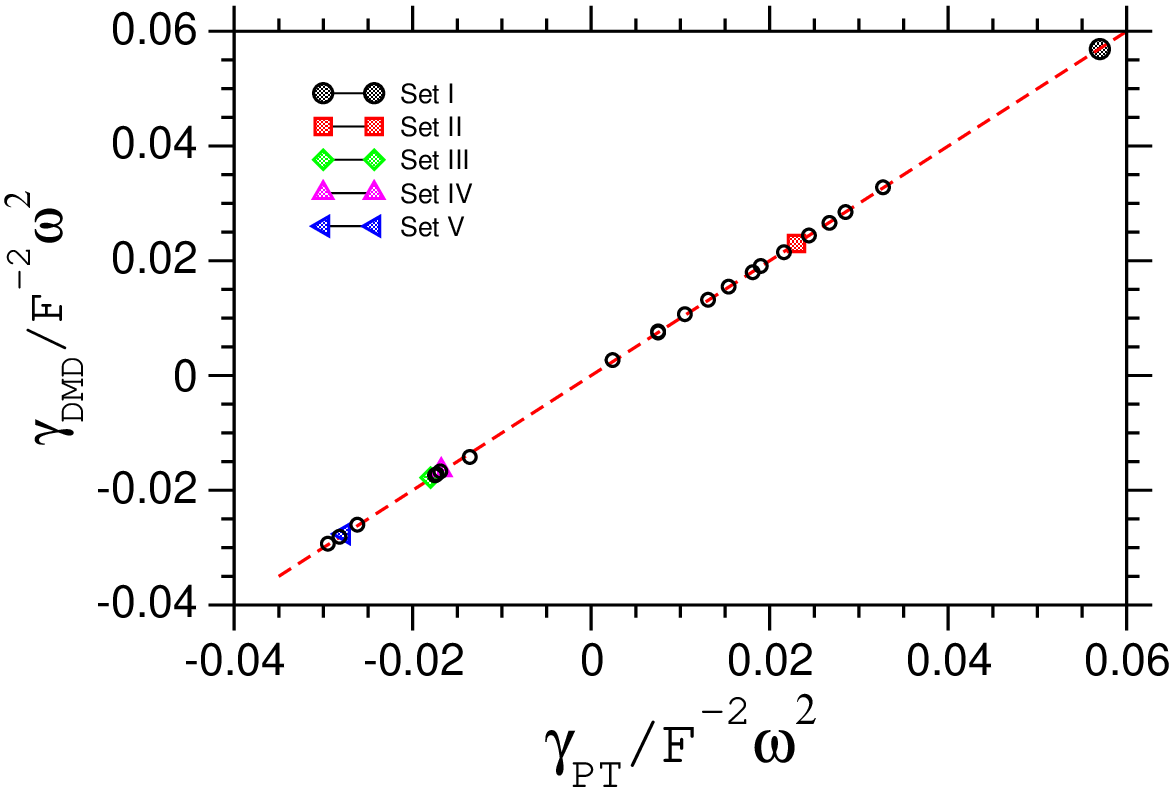}
  \caption{$\gamma_{\textrm{DMD}}$ versus $\gamma_{\textrm{PT}}$ for the five sets of parameters in the case of 5-modes coupled systems. The red dashed line corresponds to $\gamma_ {\textrm{DMD}}=\gamma_{\textrm{PT}}$. Filled colored symbols correspond to the five specific simulations of figure \ref{fig:fig4}.} 
 \label{fig:fig5}
\end{figure}

More systematically, we have determined all anharmonicity parameters for the 5 modes of the 5 systems, and Fig.~\ref{fig:fig5} compares the values obtained from the DMD approach with the perturbation theory prediction.
The agreement is remarkable, for a broad range of anharmonicities both negative and positive, with a Pearson correlation coefficient of 0.9999 and an average error of 1.31 \% between the two methods.



\subsection{Cubane molecule}
\label{ssec:cubane}

We finally consider a realistic molecular system with the case of the cubane molecule C$_8$H$_8$, whose dynamics is described by a quantum tight-binding (TB) model taken from Ref. \cite{VO2002}. For this system, all 42 vibrational modes are coupled to each other, but owing to the very high symmetry of cubane ($O_h$ point group) only 3 modes are infrared active. The details of the TB Hamiltonian, which only considers the valence 1s orbitals of hydrogen atoms and 2s, 2p orbitals of carbon atoms, are given in the original reference \cite{VO2002}. Here, we also use a more realistic external excitation under the form of a radiation field $\vect{E}(t)=E_0\bm{\mu}\sin(\Omega t)$, with $E_0$ its magnitude and ${\bf u}$ a fixed unit vector. This field couples to the molecule in the electric dipolar approximation, such that $H_c({\bf q},t,\Omega)=-\bm{\mu} ({\bf q})\cdot \vect{E}(t)$ where $\bm{\mu}({\bf q})$ is the dipole moment vector \cite{calvojpc14}.

Within the present TB model, the three IR active harmonic frequencies of cubane are found at 923.2, 1345.6, and 3064.5~cm$^{-1}$, with relative IR harmonic intensities in the ratios 0.276:1:0.001. For this system, perturbation theory introduces an additional approximation to the true potential energy surface, and we resort to an alternative method to determine the anharmonic coefficients of the three active modes, which is in principle also more accurate and in closer connection with experiments.

More precisely, accurate anharmonic vibrational spectra were first determined using the standard approach based on the Fourier transform of the electric dipole time autocorrelation function, time series of the electric dipole vector being collected in constant-energy MD trajectories conducted without external field, but initiated from a sample at a prescribed temperature $T$. Various spectra were then obtained between 100~K and 1000~K in steps of 100~K, from which the temperature-dependent anharmonic frequencies $\widetilde\omega_i(T)$ were deduced.

The anharmonicity coefficients $\gamma_i$ we seek are defined from the energy dependence of $\omega_i$, and the connection between this energy dependence and the temperature dependence is provided by the knowledge of the microcanonical density of states $W(E)$, from which 
\begin{equation}
\widetilde\omega_i (T) = \frac{\int_0^\infty \widetilde\omega_i(E) \, W(E) \, e^{-\beta E} \,  dE}{\int_0^\infty W(E) \, e^{-\beta E} \, dE},
\end{equation}
where $\beta=1/k_{\rm B} T$. Inserting the defining relation $\widetilde\omega_i(E)=\omega_i(1+\gamma_i \times E)$, the anharmonicity coefficient can be obtained from
\begin{equation}
     \gamma_i = \frac{ \widetilde\omega(T) - \omega_i}{  \omega_i \times \langle E \rangle(T)}
     \label{eq:omegaET}
\end{equation}
with the thermally averaged energy $\langle E\rangle$ given by
\begin{equation*}
\langle E\rangle (T) = \frac{\int_0^\infty E W(E) \, e^{-\beta E} \,  dE}{\int_0^\infty W(E) \, e^{-\beta E} \, dE}.
\end{equation*}
Note that the latter quantity can also be obtained directly as a time average from the MD simulations in the canonical ensemble used to sample the initial conditions for the microcanonical trajectories generating the time series of the dipole moment vector. Hence there is no need to determine the microcanonical density of states to evaluate the anharmonic coefficient from the canonical data. We refer to $\gamma_{\rm MD}$ the value of the anharmonic coefficient so inferred from the canonical trajectories and the analysis of the temperature-dependent IR absorption spectra, also noting that the intramode and intermode contributions cannot be obtained separately using this method.

In addition to this reference approach, second-order perturbation theory can also be applied, using numerical third and fourth derivatives of the TB potential energy at the equilibrium geometry. The PT method provides the intermode and intramode contributions to the global anharmonicity for each mode, which we denote as $\gamma_{\rm PT}$ possibly with the superscripts (inter) or (intra) accordingly. 

In applying the driven MD strategy to the present system, an electromagnetic field with magnitude $E_0=4\cdot 10^{-4}$ atomic units, oriented perpendicularly to an arbitrary facet of the cubic carbon skeleton, is used and no excess energy is given in the triply degenerate modes. 

\begin{figure}[htb]
\begin{center}
 \includegraphics[width=8.cm]{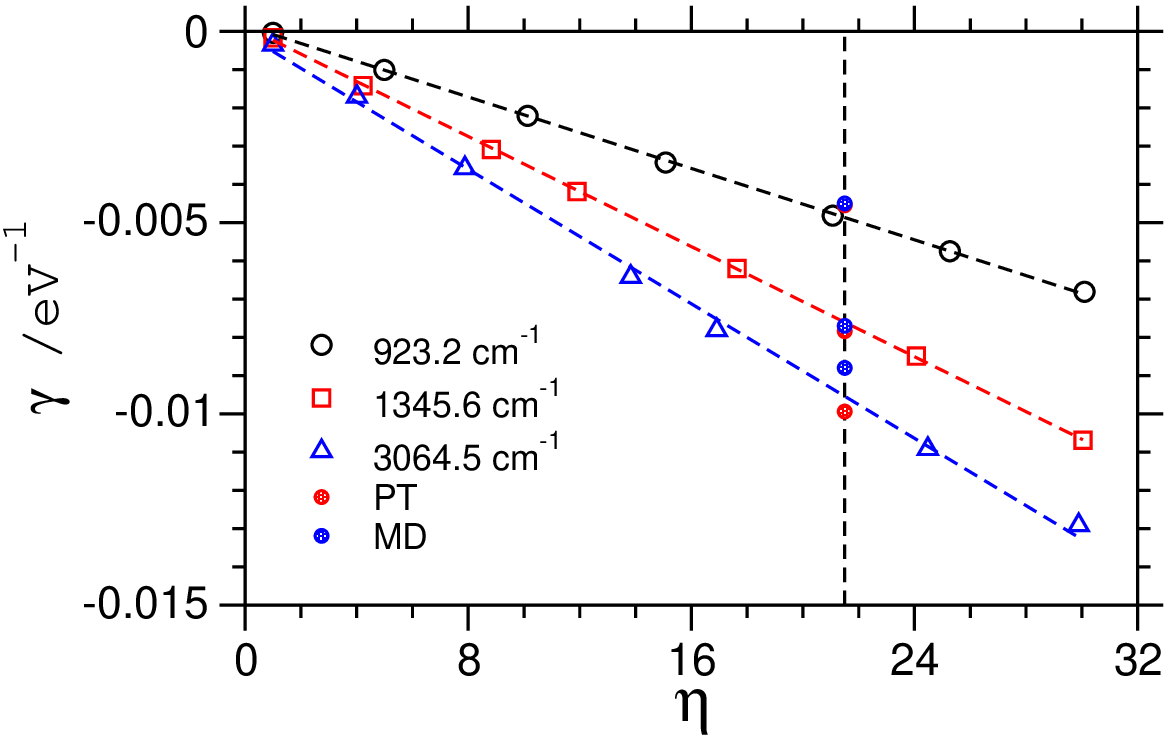}
  \caption{Effective anharmonicity coefficient $\gamma_{\rm{DMD}}$ obtained from DMD simulations as a function of the relative excess energy $\eta$ for the three IR active modes of the cubane molecule. The dashed lines passing through the points correspond to linear regressions, while the vertical dashed line highlights the intersection at $\eta=(g+1)/2=21.5$ where the true anharmonicity coefficients can be read. The predictions from perturbation theory (PT) and from the analysis of the temperature-dependent IR spectra (MD) are also indicated.}  
  \label{fig:fig8}
\end{center}
\end{figure}

The excess energy was chosen to cover the relative range $\eta=0$--32, and the DMD trajectories themselves were propagated over about 200 vibrational periods for each mode, or a few picoseconds only, using a time step of 0.03~fs. The variations of the effective total anharmonicity coefficient $\gamma$ with increasing $\eta$ are represented in Fig.~\ref{fig:fig8} for the three IR active modes.
The results shown in this figure reveal that even for such a relatively large system the effective anharmonicities still behave again almost linearly with increasing deposited energy, thus allowing a clear estimation of the true anharmonicities for the three modes considered. Comparison with the two alternative methods also appears very satisfactory in Fig.~\ref{fig:fig8}. More quantitatively, the intramode and intermode contributions to $\gamma_i$ obtained from both the DMD and PT methods are detailed in Table \ref{tab:cubane}, together with the global anharmonicities also obtained from the analysis of the canonical spectra (MD method).

\begin{table}
\begin{center}
\begin{tabular}{c|ccc|ccc|c}
\hline \hline
Harmonic & \multicolumn{3}{c|}{DMD} & \multicolumn{3}{c|}{PT} & MD \\
frequency & intra & inter & total & intra & inter & total
& total \\
      \hline
   923.2     &   -0.04   &-4.88 & -4.92 & -0.05 & -4.50 & -4.55 &  -4.5 $\pm$ 0.2\\
    1345.6  &  -0.17 & -7.40 &   -7.57 & -0.20 & -7.63   &-7.84 &  -7.7 $\pm$ 0.2\\
     3064.5     &     -0.35   &-9.42 &  -9.77 & -0.80  & -9.14   &-9.94  &  -8.8 $\pm$ 0.2\\
      \hline\hline
\end{tabular}
\caption{Anharmonicity coefficients for the three active IR vibrational modes of the cubane molecule obtained from the DMD, PT and MD approaches. The harmonic frequency is given in cm$^{-1}$, anharmonicities are given in units of $10^{-3}$~eV$^{-1}$.}
\label{tab:cubane}
\end{center}
\end{table}

The performance of the DMD method is particularly remarkable for the three modes, the most significant discrepancy being found for the C-H stretching mode at 3064.5~cm$^{-1}$, and especially its intramode contribution. For this mode the direct MD evaluation from the spectral analysis seems to indicate a lower anharmonicity (in magnitude) relative to both the PT and DMD approaches, which could be due to some contributions from higher order terms (beyond quartic), or to some mode mixing in the IR spectra, involving especially the nearby other C-H stretching modes at 3051, 3054, and 3085~cm$^{-1}$ and their possible IR activation due to anharmonicities, an effect that was neglected from the perturbative treatment. Anharmonicities of the electric dipole moment surface, or those of the potential energy surface beyond second order, are not included in the PT approach but could become significant in the temperature range of 100--1000~K at which the IR absorption spectra were calculated.

\section{Concluding remarks}
\label{sec:ccl}

Driven molecular dynamics was originally introduced to evaluate vibrational frequencies of large dimensional systems as a computationally effective method alternative to eigenvalue determination \cite{Bowman2003}. Following a similar idea, we have extended the method for calculating anharmonicity coefficients in a perturbative perspective, assuming the harmonic solution to be known. The gain in computational efficiency is similar to the harmonic case: a traditional anharmonic calculation based on vibrational perturbation theory requires derivatives of the potential energy surface up to order 4, leading to a scaling with size exceeding $N^3$. With driven MD, the numerous derivatives are replaced by simple force evaluations (requiring the normal gradient), but only limited integration of the trajectory over some hundreds of vibrational cycles per mode of interest, or just a few picoseconds in practical cases of molecular systems. While we have assumed the harmonic frequencies to be known from the start, DMD could also be used to determine them in practice, before its application to the anharmonic case. This would be particularly useful for large systems for which the diagonalization of the Hessian matrix is involved.

Another practical advantage of evaluating anharmonicities from DMD simulations rather than PT should be in the context where the potential energy surface uses an explicit description of electronic structure, be it with DFT or any {\em ab initio} method. The higher energy derivatives would typically be determined by numerical differentiation of (at best) the analytical Hessian. For large systems, small errors in the electronic energy can induce large changes in the successive derivatives, compromising the robustness of perturbation theory. In contrast, both the MD and DMD approaches are expected to be potentially more stable.

Having provided rigorous grounds in one dimension and generalized the algorithm for coupled oscillators, we have applied the method to a realistic molecular system showing only very few active infrared modes owing to its very high symmetry. Such a situation is particularly appealing for our approach, since the anharmonicities can be extracted individually, reducing accordingly the overall load. In contrast, traditional perturbation theory requires all the couplings between the mode of interest and all other modes to be determined individually. In this respect, other carbonaceous molecules such as fullerenes or polyaromatic could be very interesting candidates to apply the present framework, especially in the light of their experimental relevance in astrochemistry where anharmonicities themselves could be characterized \cite{joblin95}. In the context of experimental comparison, the present method could be used to infer the temperature-dependent anharmonic coefficients, rather than their energy-dependent microcanonical counterpart, by simple application of Eq.~(\ref{eq:omegaET}).

However, the method can also be straightforwardly extended to extract the entire Dunham matrix. From there, temperature and even kinetic effects over macroscopically long times could be taken into account through dedicated Monte Carlo schemes \cite{cpbasire13}. Additionally, it would be possible to incorporate nuclear quantum effects, still at a perturbative level and even though all anharmonic coefficients were evaluated from a purely classical framework. This would be particularly useful at low temperatures or for high-frequency modes involving light atoms.  We plan to explore these ideas further in future work.

\begin{acknowledgements}
The authors gratefully acknowledge financial support by the Agence Nationale de la Recherche (ANR) Grant No. ANR-16-CE29-0025, and the GDR EMIE 3533.
\end{acknowledgements}

\bibliographystyle{spphys}
\bibliography{biblio}
  
\end{document}